\documentclass[conference]{IEEEtran}
\IEEEoverridecommandlockouts

\usepackage{cite}
\usepackage{amsmath,amssymb,amsfonts}
\usepackage{algorithmic}
\usepackage{graphicx}
\usepackage{textcomp}
\usepackage{xcolor}
\usepackage{xspace}

\newcommand{\pg}{PostgreSQL\xspace }
\newcommand{\job}{JOB\xspace}

\def\BibTeX{{\rm B\kern-.05em{\sc i\kern-.025em b}\kern-.08em
    T\kern-.1667em\lower.7ex\hbox{E}\kern-.125emX}}
\begin{document}

\title{How I Learned to Stop Worrying and Love Re-optimization
}

\author{\IEEEauthorblockN{Matthew Perron}
\IEEEauthorblockA{MIT CSAIL \\
mperron@csail.mit.edu}
\and
\IEEEauthorblockN{Zeyuan Shang}
\IEEEauthorblockA{MIT CSAIL \\
zeyuans@mit.edu}
\and
\IEEEauthorblockN{Tim Kraska}
\IEEEauthorblockA{MIT CSAIL \\
kraska@csail.mit.edu}
\and
\IEEEauthorblockN{Michael Stonebraker}
\IEEEauthorblockA{MIT CSAIL \\
stonebraker@csail.mit.edu}
}

\maketitle

\begin{abstract}
Cost-based query optimizers remain one of the most important components of database management 
systems for analytic workloads. Though modern optimizers 
select plans close to optimal performance
in the common case, a small number of queries are an order of magnitude slower than they 
could be. In this paper we investigate why this is still the case, despite decades of improvements 
to cost models, plan enumeration, and cardinality estimation. We demonstrate why we believe 
that a re-optimization mechanism is likely the most 
cost-effective way to improve end-to-end query performance. We find that even a simple 
re-optimization scheme can improve the latency of many poorly performing queries. We demonstrate that re-optimization improves the end-to-end latency of the top 20 longest running queries in the Join Order Benchmark by 27\%, realizing most of the benefit of perfect cardinality estimation.
\end{abstract}


\section{Introduction}
\label{sec:intro}

The basic structure of query optimizers dates to the pioneering paper in 
1979~\cite{selinger1979access} and was adopted by essentially all commercial products. A cost model for all steps of a query plan, methods for estimating the selectivity of all possible intermediate 
tables, and a dynamic programming algorithm to search the space of possible plans were introduced. Many 
enhancements have been added over the years, for example, inclusion of bushy search strategies, 
early termination, better statistics, and cost models. 

However, this 
class of optimizers is known to suffer serious performance problems. Notably, selectivity 
estimates assume the independence of columns in a table or across joins and also assume a uniform 
distribution of data elements in each column. In the real world, correlation between columns is 
widespread. For example, salary is usually correlated with age. Also, column values are 
often strongly skewed, for example 40 stocks out of 4000 in the NYSE account for 50\% of the total 
volume. Moreover, estimates for the size of intermediate tables become increasingly imprecise as 
one ascends higher in the plan tree.

Surprisingly, even with simplifying assumptions, query optimizers mostly choose plans that are near optimal performance even when cardinality estimates are wrong. But in a minority of cases, the optimizer chooses plans that are many times slower than optimal. This can significantly slow down the end-to-end execution time of query workloads. Put simply, a small number of optimization mistakes leads to workload performance far below what is possible. 
Ideas to improve the optimizer in these cases include better selectivity estimates~\cite{estan2006end, chen2017two, leis2017cardinality} providing more exotic search mechanisms for plans~\cite{marcus2018deep, ortiz2018learning}, 
and changing query plans that have gone “off the rails” at runtime~\cite{kabra1998efficient, babu2005proactive, kaftan2018cuttlefish, avnur2000eddies, deshpande2007adaptive}. A more recent trend is to recast query optimization as a machine learning problem. The purpose of this paper is to 
shed light on profitable directions to explore.

\begin{figure}[t]
    \centering
    \includegraphics[width=\columnwidth]{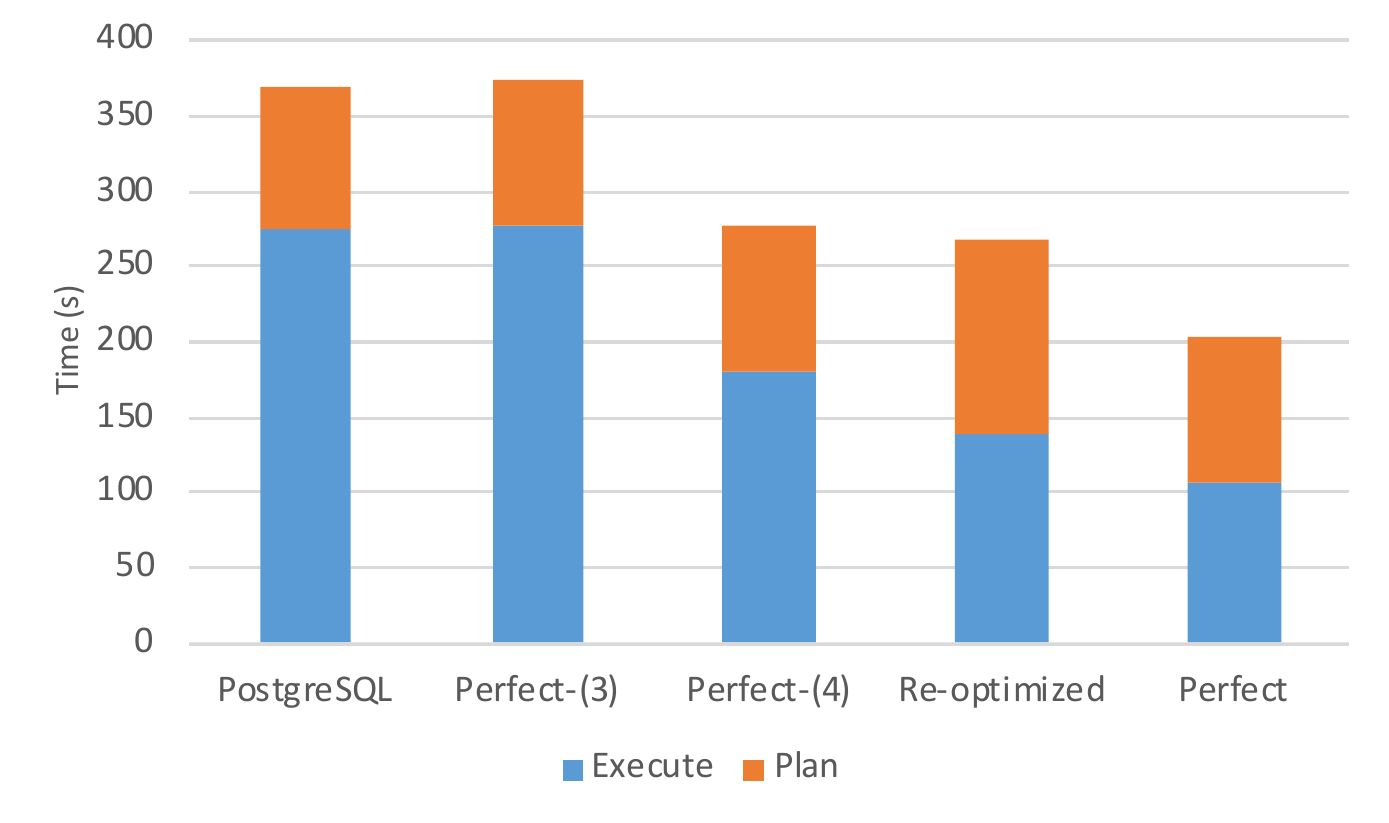}

    \caption{
        Total Query planning and execution times for the top 20 longest running queries in \job with the default \pg optimizer.
    }
    \label{fig:top_20_comparison}
\end{figure}

We consider the Join Order Benchmark (\job) proposed by Leis et al.~\cite{leis2015good}. \job is a workload of 113 queries, with varying numbers of joins. It is considered more challenging for query optimizers than standard decision support benchmarks. Our target workload is composed of queries from applications with the assumption that plans are cached. While most queries in the benchmark have execution times close to those produced with perfect estimates, we find that just 20 sub-optimal queries slow execution time of the benchmark by a factor of two, not including query planning time. 

Leis et al. provide experimental evidence that cardinality estimates, particularly join cardinality estimates, are poor and can result in queries with noticeably 
sub-optimal plans. The 
inference is that improving estimates would result in better plans. 
In Figure~\ref{fig:top_20_comparison} we consider the 20 longest running queries in \job without cached plans. We define perfect-(n) as a cardinality estimator with perfect estimates for joins of n tables and fewer. We then compare the \pg optimizer with perfect-(3), perfect-(4), a re-optimization scheme, and all perfect estimates. Perfect-(3) achieves no improvement for these queries, while Perfect-(4) and re-optimization improve end-to-end latency by 25\%. This is a worst-case scenario since we include planning time. If we assume plans are cached, the improvement to execution time is over 35\%, including about 30 seconds to re-optimize. From this experiment we can conclude that improvements to cardinality estimates that do not approach perfect-(4) will not improve the end-to-end latency of the benchmark. With proposed techniques this level of improvement seems out of reach. However, we also find that much of the benefit of perfect estimates can be achieved with re-optimization strategies. 

In the remainder of this paper we first discuss the query optimizer architecture, and the prospects of improving each of its components. Given the work of Leis et al.~\cite{leis2015good} and our own study, we evaluate the likelihood that these improvements will help in Section~\ref{sec:background}.

In Section~\ref{sec:job}, we confirm this but demonstrate that perfect estimates on all joins of four tables and fewer are required before a marked improvement to the workload execution time is achieved. We conclude that such a substantial improvement seems unlikely. In Section~\ref{sec:join_card_est_difficulty}, we discuss the causes of cardinality estimation errors and how they cause poorly performing plans, using examples of \job queries. 

In Section~\ref{sec:reopt} we consider re-optimization as an alternative to improving cardinality estimates. We simulate a simple re-optimization scheme and find that it achieves much of the benefit of perfect cardinality estimates with minimal changes to the optimizer. We then discuss risks of re-optimization schemes.

Finally, we conclude with a discussion of future work in Section~\ref{sec:conclusion}.

\section{Background}
\label{sec:background}

Work on query optimization typically focuses on cost models~\cite{babcock2005towards, manegold2002generic}, cardinality estimation~\cite{kipf2018learned, leis2017cardinality} and 
plan enumeration~\cite{marcus2018deep,ortiz2018learning,krishnan2018learning}. Below we discuss 
each of these in turn, and their prospects for improving workload execution time.

\subsection{Cost Models}
\label{sec:cost_models}

Cost models for commercial query optimizers are notoriously difficult to understand and tune. Cost 
functions have parameters that depend on the properties of the machine on which the DBMS is running as well the workload. For example, \pg includes parameters for the CPU cost of processing a 
tuple and the cost of reading a random page from disk, both system dependent. It also includes a 
parameter for the ``\texttt{effective\_cache\_size}'' dependent on the likelihood of concurrent queries evicting tuples from the cache. \pg documentation even warns against changing constants without strong evidence. Commercial systems have similarly complex and difficult to tune cost models.  
There have been several studies on cost models. Babcock et al. consider the probability distributions of cost \cite{babcock2005towards}, and \cite{manegold2002generic} proposes a generic cost model for hierarchy memory database systems.

While it might be possible to improve a cost 
model using a machine learning approach, or by benchmarking the system to determine appropriate 
constants, evidence suggests that cardinality estimates play a much larger role in poor query 
performance than inaccurate cost models~\cite{leis2015good}. Since cardinality 
estimates are an input to the cost function, when estimates are orders of magnitude off, the model has no hope of predicting cost correctly. In summary, while complex and inaccurate cost functions are a problem, fixing them 
is unlikely 
to improve query optimizers without improving cardinality estimates first. 

\begin{table}
    \centering
    {\scriptsize {
\setlength{\tabcolsep}{3pt}
\begin{tabular}{l|l}
	\hline
        \textbf{\# Tables in Join} & \textbf{\# Estimates} \\
        \hline
    1 & 977 \\
    2 & 1346 \\
    3 & 2676 \\
    4 & 4493 \\
    5 & 6510 \\
    6 & 8387 \\
    7 & 9781 \\
    8 & 10326 \\
    9 & 9732 \\
    10 & 8019 \\
    11 & 5665 \\
    12 & 3357 \\
    13 & 1630 \\
    14 & 624 \\
    15 & 177 \\
    16 & 33 \\
    17 & 3 \\
\end{tabular}
}
}
    
    \vspace{5px}
    \caption{
        Number of cardinality estimates on joins of n tables in \job
    }
    \label{tab:job_num_cardinalities}
\end{table}

\subsection{Plan Enumeration}
\label{subsec:plan_enumeration}

For each query the query optimizer explores the space of possible plans to find one with relatively low cost. Finding the lowest cost plan is known to be NP-complete 
~\cite{ibaraki1984optimal}. Because of extreme computational complexity, the 
optimizer usually explores a subset of possible plans. For instance, 
System R 
used a dynamic programming approach and explores all left-deep plans with the exception of Cartesian products~\cite{selinger1979access}. 
Modern optimizers often use dynamic programming but allow bushy plans and explore the search space 
using a set of heuristics~\cite{graefe1993volcano, chaudhuri1998overview, graefe1995cascades, chen2016memsql}. Many other papers improve on the original model with a 
variety of 
methods including genetic algorithms~\cite{postgres, bennett1991genetic}, randomization~\cite{waas2000join}, and deep 
learning~\cite{krishnan2018learning,ortiz2018learning,marcus2018deep}. Each of these strategies improves query 
optimization by exploring plans not explored by other methods, or by finding a low-cost plan faster 
than other strategies. 

However, all of these strategies use the same cost models described in Section 
\ref{sec:cost_models} to determine the best plans. While improvements to plan enumeration may allow 
the optimizer to choose a different set of plans, or to choose a good plan faster, improvements to 
query execution times will likely remain elusive as long as the costs associated with those plans 
continue to be off due to huge cardinality estimation errors. In experiments with \pg, we 
found that a good plan was in the set of plans searched, but a bad plan's cost was estimated to be 
much lower than this good plan. This is obvious given the performance gains we achieve with perfect estimates in Figure~\ref{fig:top_20_comparison}.  With perfect estimates the workload execution time is half of that with \pg estimates. Therefore we believe that improvements to plan enumeration strategies alone are unlikely to achieve significant performance improvements.

\subsection{Cardinality Estimation and Statistics}
\label{subsec:card_est}
 To choose a plan, the optimizer makes many cardinality estimates. Depending on the join graph and the number of relations, the number of estimates could be in the thousands for a single query. The standard method of estimating cardinalities is to scan all the tables in the database (or a 
significant sample of each table), and build statistics on each \cite{ioannidis1993optimal, ioannidis1995balancing, poosala1996improved, jagadish1998optimal}. These statistics 
most often include one 
dimensional equi-depth or equi-width histograms on each column in a table, a list of most frequent 
values and their frequencies, the number of distinct values, and min and max values~\cite{piatetsky1984accurate}. Some commercial systems allow a DBA to create column group statistics on a subset of columns in order to capture a correlation. A selectivity estimate of each 
predicate is then made using the available statistics on each columns and a set of simplifying assumptions including independence of predicates and uniform distribution of values. 

But with these strong assumptions join cardinality estimation errors increase exponentially with the 
number of joins~\cite{ioannidis1991propagation}. This is problematic as many estimates critical for choosing a good plan are on joins of many tables. Table~\ref{tab:job_num_cardinalities} shows the number of cardinalities for multi-way joins in all queries in \job. The vast majority of estimates are for joins.

Many papers have been written on how to improve cardinality estimates. Techniques include sampling 
~\cite{leis2017cardinality,chen2017two,estan2006end}, improved histograms~\cite{jagadish1998optimal}, 
sketches~\cite{alon2002tracking }, correcting for correlations between columns 
or 
across joins~\cite{ilyas2004cords}, and deep learning~\cite{kipf2018learned}.  

But, while many improvements to cardinality estimation have been suggested, only a tiny fraction have 
been implemented in widely used DBMSs. We hypothesize the reasons for this are two-fold. First, 
these methods add another layer of complexity to an already very complex system, sometimes require 
additional table scans or are slow to build or make predictions.  As we demonstrate in Section~\ref{sec:job}, even with the many simplifying assumptions in the query optimizer, it still produces 
good plans in most cases. Only a subset of queries have performance more than 2 times worse than plans produced with perfect cardinalities. Furthermore, improving the average cardinality estimation error is unlikely to help. As long as a single poor plan in the search space exists with lower cost than a good plan due to underestimated cardinalities, execution time may suffer. 

\subsection{Query Re-optimization}
\label{subsec:reopt}
Unlike the techniques described above, Query Re-optimization does not necessarily involve 
significant changes to the query optimizer itself. Instead, re-optimization detects when a 
query plan has behavior sufficiently different from what was predicted and attempts to correct the plan after execution begins. 
Query re-optimization is a tacit admission that the query optimizer can make poor 
decisions and attempts to mitigate them. Because the optimizer often chooses good plans, as we see in Table~\ref{tab:query_runtime_distribution}. The hope is for re-optimization to allow the system to improve the worst performing queries without slowing down well planned queries.

Dynamic mid-query re-optimization was first proposed and implemented by Kabra et al. in 1998~\cite{kabra1998efficient}.
Query re-optimization add statistics collection operators to the query plan. When the distribution of data passing through the operator deviates too much from what is 
expected, the 
system re-plans the remainder of query in light of this new information. The optimizer  can then choose different join orders, access paths, or join algorithms.

Adaptive query processing is another technique that changes the execution plan of the query at runtime. Many of these techniques cannot optimize join orders but instead can choose an implementation of an operator at runtime. Cuttlefish, for example cannot change the join order, but chooses operator implementations at runtime with a reinforcement learning approach~\cite{kaftan2018cuttlefish}. Work like Eddies~\cite{avnur2000eddies} can change the join order at runtime but only for ripple joins~\cite{haas1999ripple}. 

\section{Another Look at the Join Order Benchmark}
\label{sec:job}

In this section we execute the \job, measuring query planning and execution time while changing the cardinality estimates that \pg uses to generate costs.

\subsection{Experimental Setup}
\label{subsec:setup}

We use a version of \pg 10.1 that we modified to allow us to replace the \pg  cardinality estimates
with arbitrary values. We then compare the performance of different cardinality 
estimation schemes. We add foreign key indexes making access path selection more challenging. 

We use the Join Order Benchmark (JOB) proposed 
by Leis et al.~\cite{leis2015good}. \job contains a set of 113 queries on the 
Internet Movie Database (IMDB) dataset. This real world dataset includes both skew and correlation 
making it difficult for query optimizers to choose efficient join orders~\cite{leis2015good}. Each 
query is a select-project-join query with 4 to 17 tables with equi-joins only. The distribution of the number of tables in queries is available in Table~\ref{tab:job_num_tables_distribution}.  Queries 
in \job have proved more difficult for optimizers to choose good plans than standard decision 
support benchmarks like TPC-H~\cite{tpc-h}. As a result, \job is a popular choice for evaluation in recent query optimization work.

\begin{figure}[t]
    \centering     
    \includegraphics[width=\columnwidth]{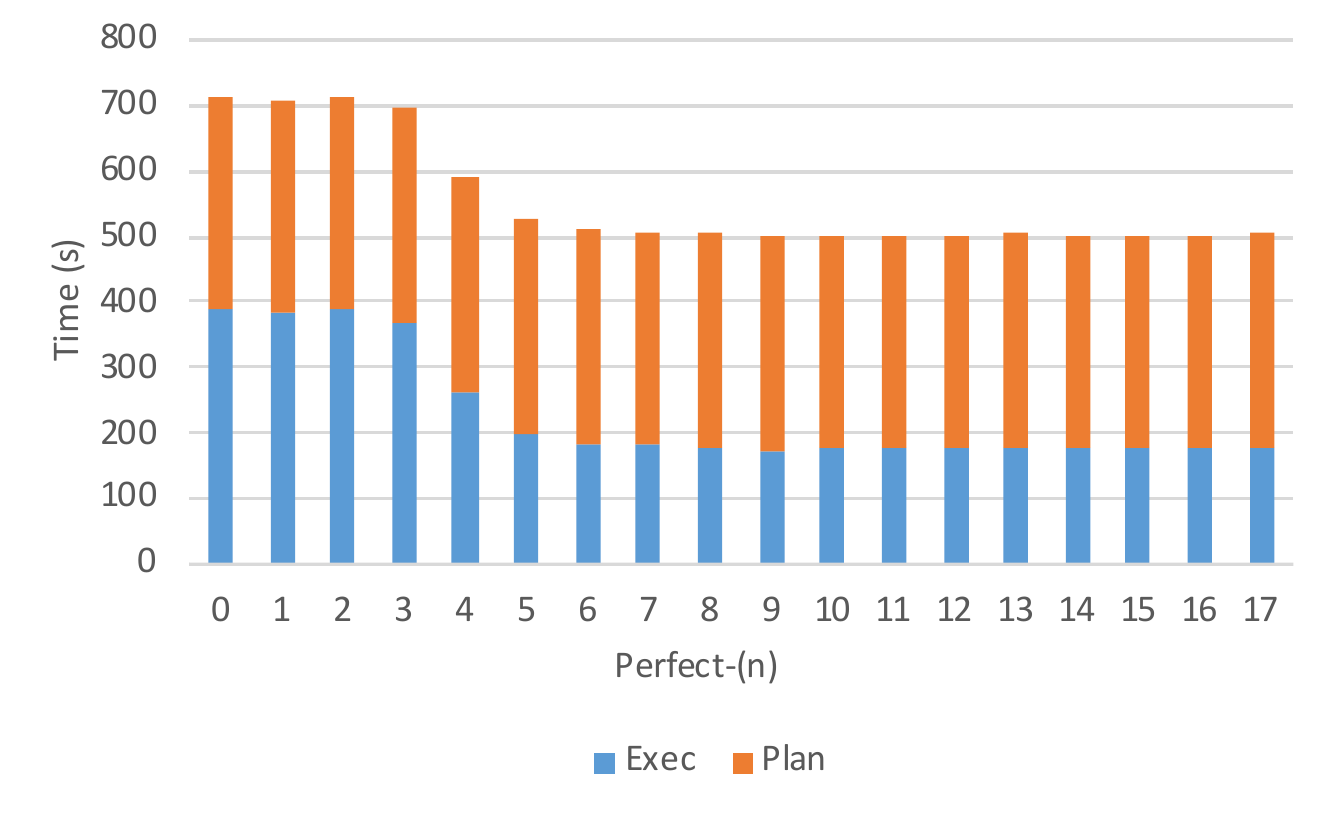}
    \caption{
        Total execution and planning time of all queries in \job with varying cardinality estimation quality
    }
    \label{fig:perfect_est_comparison}
\end{figure}
\begin{table}
    \centering
    {\scriptsize {
\setlength{\tabcolsep}{3pt}
\begin{tabular}{l|l}
	\hline
        \textbf{Relative Runtime} & \textbf{Number of Queries} \\
        \hline
    0.1 - 0.8            &  7 \\
    0.8 - 1.2            &  32 \\
    1.2 - 2.0            &  28 \\
    2.0 - 5.0            &  32 \\
    $>$ 5.0            &  14 \\
\end{tabular}
}
}
    \vspace{5px}
    \caption{
        Execution time of \job queries with \pg cardinality estimation relative to perfect-(17)
    }
    \label{tab:query_runtime_distribution}
\end{table}

We execute our experiments on  a \texttt{n1-highmem-4} virtual machine instance on Google Cloud 
Platform (GCP) with a 
256GB SSD persistent disk, and 26 GB of memory. We disable intra-query parallelism, and set 
\texttt{shared\_buffers}, 
\texttt{temp\_buffers}, and \texttt{work\_mem} to 400MB each. Because \pg, in addition to managing it's own buffer pool, uses the file system's large buffer cache, all tables and indexes are cached in memory.

To give \pg the best chance at good cardinality estimates, we set the \texttt{default\_statistics\_target} to it's maximum value, 10,000, and execute \texttt{ANALYZE} to collect statistics. 

We define ``planning time'' to be the time to parse and optimize the query, including time spent on re-optimization (see Section~\ref{sec:reopt}) and ``execution time'' to be the time spent on execution of query plans.

\subsection{Quality of Cardinality Estimates and Execution Time}
\label{subsec:card_est_and_latency}

We explore how the quality of improved cardinality estimates impacts the execution time of the workload. To get a rough estimate, we define the construct of perfect-(n) as a version of \pg where the cardinality estimator is given an oracle for cardinality estimates on joins of n tables and fewer. This means that perfect-(n) has an oracle for a subset of cardinality estimates of perfect-$(n+1)$. To estimate the cardinality of joins of more than $n$ tables, the default \pg cardinality estimate is used. For example, to make an estimate of a join of 5 tables in perfect-(4), the cardinality estimator receives as input perfect base table cardinalities and join cardinalities of up to 4 tables, but otherwise uses its default estimation techniques including independence and uniformity assumptions. The quality of estimates for joins of 5 tables, are, on average better with perfect-(4) than perfect-(3), etc. Perfect-(1) gives only perfect base table cardinality estimates. Since the there are at most 17 relations in \job  perfect-(17) has perfect cardinality estimates.

Figure~\ref{fig:perfect_est_comparison} shows the end-to-end runtime of the executing all queries in the \job with perfect-(n) for increasing values of n. We find perfect estimates on base tables, pairs of tables, and triples give virtually no benefit to benchmark execution time. Surprisingly, any method of improving cardinality estimation that does not achieve estimates better than perfect-(3) can expect to achieve nearly no improvement to benchmark execution time. We discuss why this is the case in following sections.

\section{The Steep Road to Good Join Cardinality Estimates}
\label{sec:join_card_est_difficulty}

\begin{table}
    \centering
    {\scriptsize {
\setlength{\tabcolsep}{3pt}
\begin{tabular}{l|l}
	\hline
        \textbf{\# Tables} & \textbf{\# Queries} \\
        \hline
    4 & 3 \\
    5 & 20 \\
    6 & 2 \\
    7 & 16 \\
    8 & 21 \\
    9 & 14 \\
    10 & 7 \\
    11 & 10 \\
    12 & 11 \\
    14 & 6 \\
    17 & 3 \\
\end{tabular}
}
}
    
    \vspace{5px}
    \caption{
        Number of queries in \job with a given number of tables
    }
    \label{tab:job_num_tables_distribution}
\end{table}

We look at the potential for execution time improvement in \job by
comparing plans generated with \pg cardinality estimates to plans generated with perfect cardinalities. While it is possible that better plans are not in the search space of \pg, perfect cardinality estimates improve execution time of the benchmark by a factor of two.
This indicates that much better plans are already in the search space of the optimizer but are simply not chosen. Less than 15\% of queries in the benchmark have execution time more 
than five times the optimal execution time, seen in Table~\ref{tab:query_runtime_distribution}. Errors in just 20 queries make up more 
than 95\% of the execution time difference of perfect estimates vs \pg estimates, not including query planning time. Surprisingly, the standard cardinality estimation model, assuming no correlation between columns and assuming uniformity chooses plans that have execution time within two times of a plan with perfect cardinalities in nearly 60\% of queries. This indicates that for most queries the additional cost of building more statistics or sampling to generate estimates will not decrease query execution time substantially, but may slow down planning time.

One challenge in improving cardinality estimates is how many must be accurately predicted. For a query like those in \job on $x$ relations, a naive optimizer implementation would have to predict $2^n-1$ cardinalities. Most optimizers prune some of the search space for simplicity. For instance, they typically do not generate plans with Cartesian products. Yet the number of estimates made by the cardinality estimator is still large. The \pg cardinality estimator makes more than 13 thousand estimates for the most complex queries in \job. Given this number of estimates, approaching perfect-(4) for a large range of queries seems to be a high bar. It is especially difficult because perfect-(4) requires perfect-(3), perfect-(2), and so forth. Clearly not every dataset and workload have the same properties as \job, and thus may require different levels of cardinality estimation improvement to find better plans. However, we believe that this is evidence of the challenges of standard approaches to cardinality estimation.

\subsection{Errors in Cardinality Estimation}

Leis et al. demonstrated that the quality of cardinality estimates has a large impact on query execution time. Below we review some of the sources of these errors, particularly for joins. While errors in base table cardinality estimates can occur because of the inaccuracies of histograms, or out of date statistics, here we focus on estimation errors that result from simplifying assumptions optimizers make with multiple predicates or in a join. With a complex join graph, as the number of relations in a query increases, the optimizer has to make many cardinality estimates. Ensuring that all of these estimates are correct seems a difficult task, given that the optimizer must make tens of thousands of estimates in \job, as seen in Table~\ref{tab:job_num_cardinalities}. 

\subsection{Correlation}
\label{sec:correlation}
The independence assumption made by the query optimizer leads to cardinality estimation errors when queries contain predicates on correlated columns. A simple example of this is a predicate on both age and salary in an employee database. Because older employees are likely to be paid more, by making an independence assumption, we may estimate fewer rows than the true value when queries have predicates on both columns. Attempts to reduce errors due to correlation include sampling techniques~\cite{estan2006end,chen2017two,leis2017cardinality} and improving estimates based on statistics discovered during execution~\cite{stillger2001leo}. But correlated predicates may be several edges away from each other on the join graph. This is referred to as a ``join-crossing correlation''. Work like CORDS~\cite{ilyas2004cords} can discover functional dependencies and correlations between pairs of columns and correct them by building column group statistics. This approach seems unlikely to improve execution time in \job, because correlations exist between columns that are several edges away in the join graph. 

\subsection{Skew}
\label{sec:skew}

\begin{table}

    \centering{{
\setlength{\tabcolsep}{3pt}
\begin{tabular}{|l|l|l|l|}
	\hline
        \textbf{id} & \textbf{symbol} & \textbf{company} & \textbf{...}\\
        \hline
    1 & APPL           &  Apple Inc. & ...\\
    2 & GOOG           &  Alphabet Inc. & ... \\
    3 & AAA            &  AAA Inc. & ... \\
    4 & BBB            &  BBB Inc. & ... \\
    ... & ...           &  ...  & ... \\
    n & ZZZ            &  ZZZ Inc. & ... \\
    \hline
    
\end{tabular}
}

}
    
    \vspace{5px}
    \caption{
        Companies Table
    }
    \label{tab:skew_example_companies}
    {
\setlength{\tabcolsep}{3pt}
\begin{tabular}{|l|l|l|}
	\hline
    \textbf{company\_id} & \textbf{shares} & \textbf{...} \\
    \hline
    
    1 & 200 & ...\\
    1 & 30 & ... \\
    2 & 80 & ... \\
    1 & 40 & ...\\
    2 & 2000 & ...\\
    4 & 20 & ...\\
    20 & 30 & ...\\
    2 & 40 & ...\\
    2 & 10000 & ...\\
    1 & 90 & ...\\
    1 & 50 & ...\\
    1 & 20 & ...\\
    1 & 30 & ...\\
    ... & ... & ...\\
    2 & 100 & ...\\
    \hline
    
\end{tabular}
}

    \vspace{5px}
    \caption{
        Trades Table
    }
    \label{tab:skew_example_trades}
\end{table}

Skew, especially skew in a join can lead to large cardinality estimation errors. It is trivial to fool cardinality estimates with skew across joins. Consider a database with stock trading information normalized into two tables. Tables~\ref{tab:skew_example_companies} and~\ref{tab:skew_example_trades} show an example including a \texttt{company} table containing data about symbols on the Nasdaq and a \texttt{trades} table containing information about trades of these stocks. Obviously companies like \texttt{GOOG} or \texttt{APPL} have higher trading volume than most other firms. Because of the uniformity assumption, the cardinality estimator significantly underestimates the number of rows a query like the one below will emit. This query simply fetches all trades of \texttt{APPL}. Note that if the predicate was instead on \texttt{company.id} of \texttt{APPL} the resulting cardinality would likely be predicted accurately because the join on the same column allows us to use the frequent value statistics of the \texttt{company.id} column. Neither \pg nor a commercial system we tested accurately predicted the join size of the following query on a generated dataset after collecting statistics. Detecting and correcting for skew with pairs of relations is possible, but achieving this for many tables becomes significantly more difficult.

\begin{verbatim}
    SELECT *
    FROM 
        company, 
        trades 
    WHERE 
        company.symbol = 'APPL' 
        company.id = trades.company_id; 
\end{verbatim}

\subsection{A Deep Dive into Slow \job Queries}
\label{sec:slow_queries}

We now look at two queries and demonstrate how cardinality estimation errors result in poor query execution time. Figures~\ref{fig:6d_join_graph} and~\ref{fig:18a_join_graph} show the join graphs of query 6d and 18a respectively. Dotted lines represent m-n relationships and arrows are n-1. All joins are equi-joins.  

\subsubsection{Query 6d}
\label{subsec:6d}

\begin{figure}[t]
    \centering     
    \includegraphics[width=0.8\columnwidth]{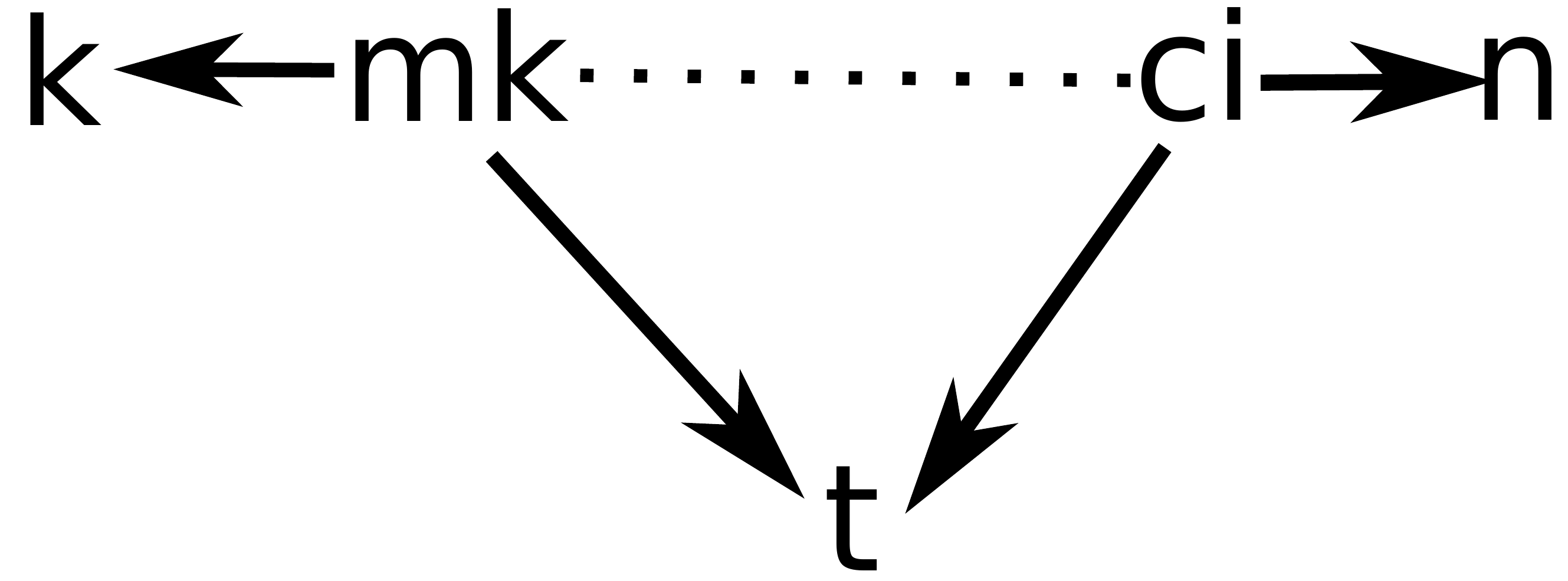}
    \caption{
        Join graph of \job query 6d
    }
    \label{fig:6d_join_graph}
\end{figure} 

Query 6d joins five tables. The join graph for this query is in Figure~\ref{fig:6d_join_graph} and the text of the query is below. 

\begin{verbatim}
SELECT min(k.keyword) AS movie_keyword,
       min(n.name) AS actor_name,
       min(t.title) AS hero_movie
FROM cast_info AS ci,
     keyword AS k,
     movie_keyword AS mk,
     name AS n,
     title AS t
WHERE k.keyword IN ('superhero',
                    'sequel',
                    'second-part',
                    'marvel-comics',
                    'based-on-comic',
                    'tv-special',
                    'fight',
                    'violence')
  AND n.name LIKE '%Downey%Robert%'
  AND t.production_year > 2000
  AND ... join conditions ...;
\end{verbatim}

The plan chosen by \pg first joins the \texttt{keyword} and \texttt{movie\_keyword} tables. The cardinality estimator correctly guesses the number of rows in the \texttt{keyword} table as 8 out of 134170 rows. Because the cardinality estimator makes a uniformity assumption and the keywords in the predicate appear frequently in the dataset, it underestimates the number of rows in the first join by more than two orders of magnitude. This is similar to the Nasdaq example discussed in Section~\ref{sec:skew}. However, correcting for this skew is not enough to fix the query. Another case of skew emerges when estimating the cardinality for the 4-way join of \texttt{cast\_info}, \texttt{keyword}, \texttt{movie\_keyword}, and \texttt{title}. This results in another underestimate of an order of magnitude, even with perfect-(2). This query does not approach optimal performance without perfect-(3).

\subsubsection{Query 18a}

\begin{figure}[t]
    \centering     
    \includegraphics[width=0.8\columnwidth]{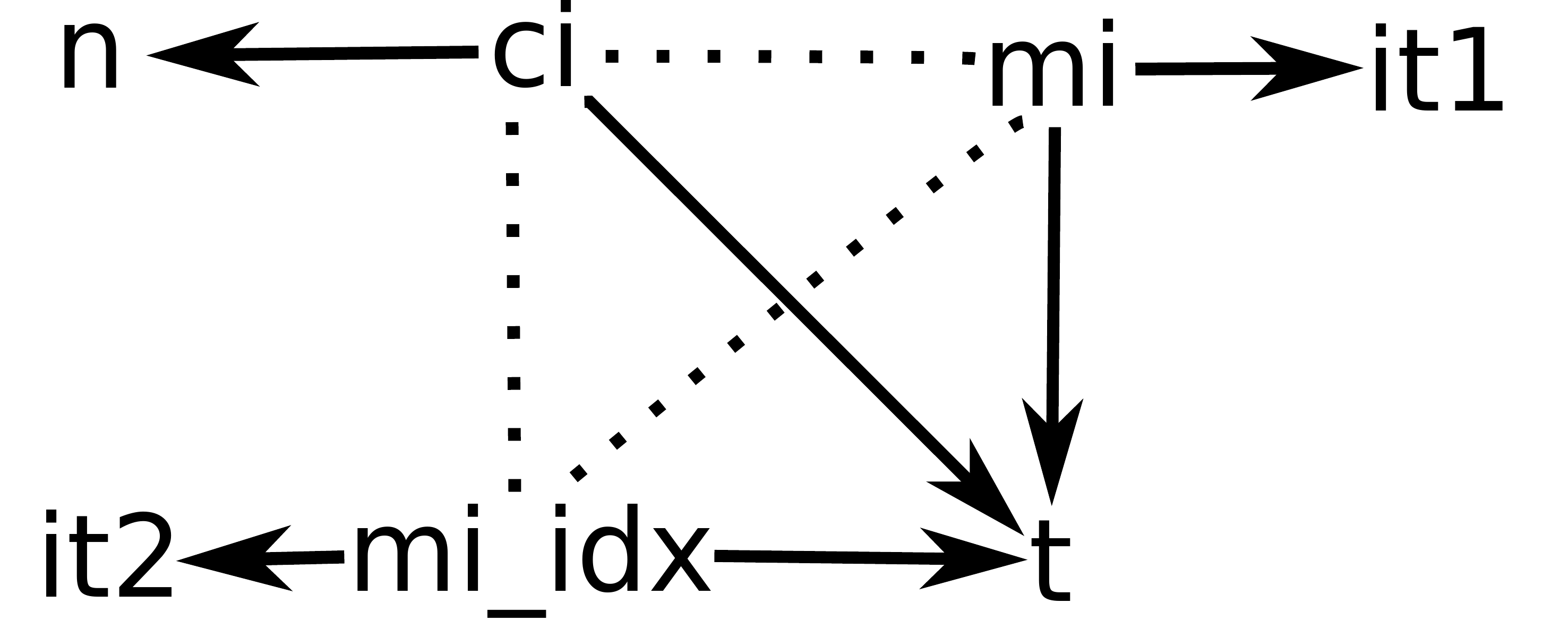}
    \caption{
        Join graph of \job query 18a
    }
    \label{fig:18a_join_graph}
\end{figure}

Query 18a has seven tables. The join graph for this query is in Figure~\ref{fig:18a_join_graph}. The text of the query is below.

\begin{verbatim}
SELECT min(mi.info) AS movie_budget,
       min(mi_idx.info) AS movie_votes,
       min(t.title) AS movie_title
FROM cast_info AS ci,
     info_type AS it1,
     info_type AS it2,
     movie_info AS mi,
     movie_info_idx AS mi_idx,
     name AS n,
     title AS t
WHERE ci.note IN ('(producer)',
                  '(executive producer)')
  AND it1.info = 'budget'
  AND it2.info = 'votes'
  AND (n.gender = 'm'
  AND n.name LIKE '%Tim%')
  AND ... join conditions ...;
\end{verbatim}

\begin{figure*}[t]
    \centering
    Query 16b
    \includegraphics[width=\textwidth]{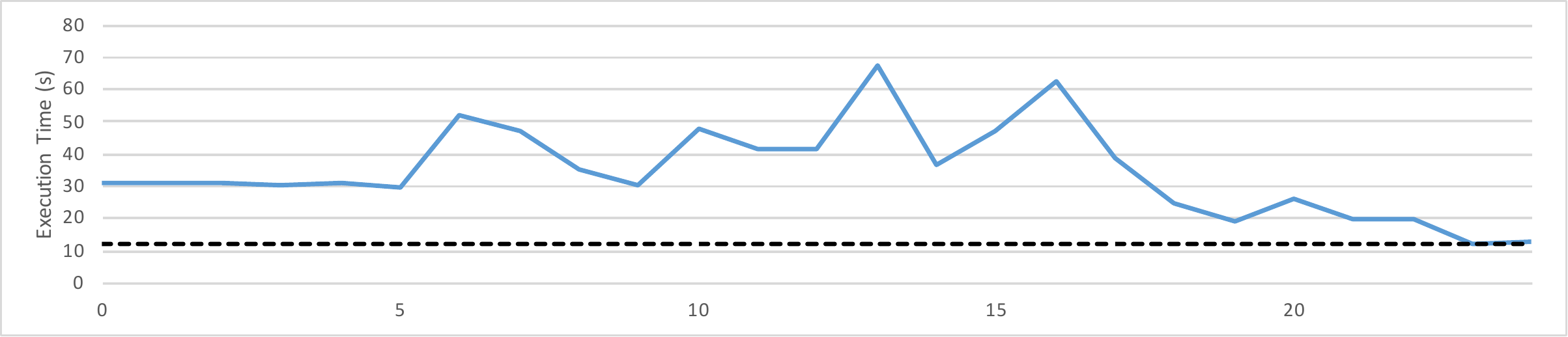}
    Query 25c
    \includegraphics[width=\textwidth]{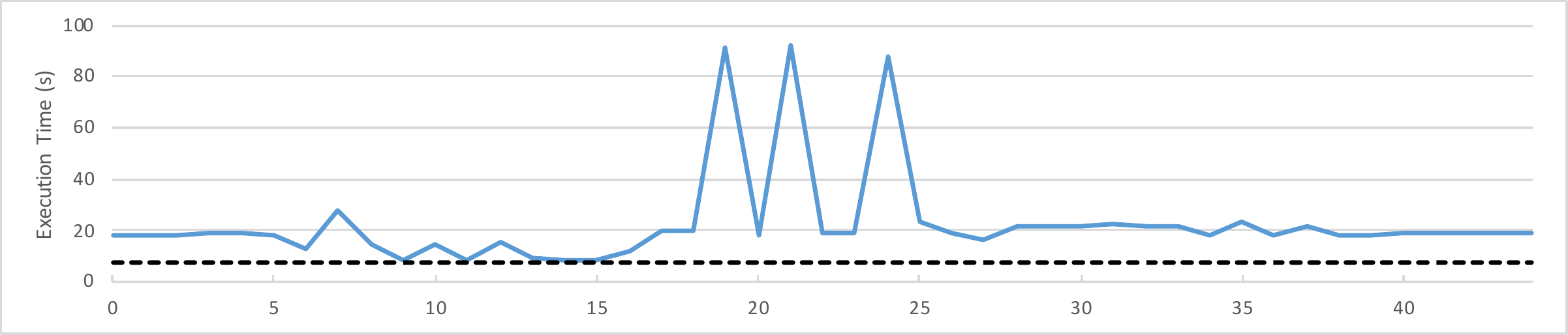}
    Query 30a
    \includegraphics[width=\textwidth]{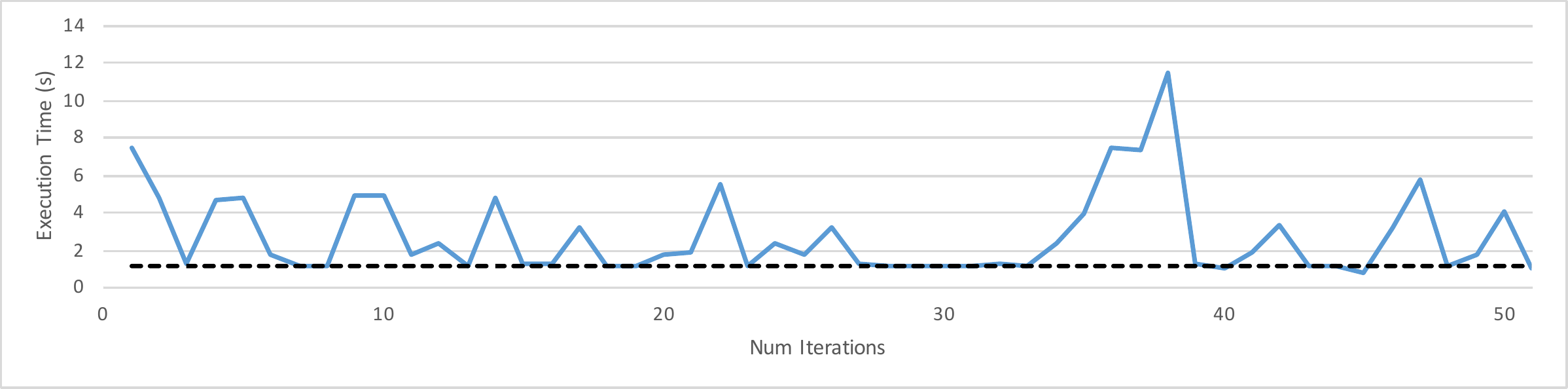}
    \caption{
        Execution time of queries with iterative improvement of cardinality estimates. The dotted line is execution time with perfect estimates. From top to bottom, plots of queries 16b, 25c, and 30a.
    }
    \label{fig:selective_improvement}
\end{figure*}

The \pg plan for this query first joins \texttt{info\_type} and \texttt{movie\_info\_idx} tables, aliased as \texttt{it2} and \texttt{mi\_idx} respectively. The cardinality estimator correctly guesses the number of rows of both base tables, but the cardinality of their join is underestimated. The estimation is 12213 rows, while the actual number is 459925. Even though the cardinality estimation is off, the actual execution time is small, so the wrong estimation does not have a significant impact on query execution time.
However, when \pg further joins these tables with \texttt{movie\_info}, the estimation error becomes more severe(estimation of 116172 rows with actual 6930334 rows), so \pg chooses a nested loop join. This join takes nearly one third of the total execution time. With perfect-(2), \pg will do a hash join instead, which is 30 times faster. Yet this does not yet approach the plan generated with perfect estimates. The cause of the slowness is the underestimation of table \texttt{it2} and \texttt{mi\_idx}.
The underestimation of this join is again a result of the independence assumption.
This query's performance only improves significantly with perfect-(4).

\subsection{Selective Improvement of Cardinality Estimates}

One suggestion to improve cardinality estimates is to discover cardinality estimation errors 
during query execution, and correct them in future executions of similar queries. LEO~\cite{stillger2001leo}, is an example of this approach. To demonstrate the limitations of these techniques, we take a set of poorly performing queries and find the lowest operator in the plan tree with cardinality estimation errors above a relative threshold of 32. We set estimates for the join and all estimates below it in the query plan to their true values and re-optimize the query. We continue this process until no operator has an estimation error larger than the threshold. This is the best case for this type of incremental improvement strategy since we execute the exactly same query repeatedly and fix cardinality estimates perfectly. In practice, correcting cardinality estimates through executed queries is more challenging since the range of queries, number of tables, is much larger than this experiment.

In 
Figure~\ref{fig:selective_improvement}, we plot the execution time of queries 16b, 25c, and 30a as we incrementally improve their cardinality estimates. We compare their execution time to that chosen using perfect cardinality estimates. For some of the worst performing 
queries, many corrections to cardinality estimates are required before discovering a good plan. Query 16b is executed 24 times before an efficient plan is chosen. While queries 25c and 30a find a good plan with only a few corrections, continuing to improve estimates actually causes the optimizer to choose plans several times slower than optimal. This means that correcting only a subset of cardinality estimates can cause the optimizer to choose query plans significantly slower than the original plan.  During plan 
enumeration, a large number of cardinalities are estimated and costs calculated. To ensure we 
arrive at a good plan, we need not only to ensure that the right plan has a reasonable cost, but 
that all terrible plans have a higher costs than good plans. Choosing a good plan may require 
correcting a large number of cardinality estimation errors, and improving a subset of estimates may cause the optimizer to choose much slower plans. 

\section{Re-optimization}
\label{sec:reopt}

\begin{figure}[t]
    \centering     
Original Query
\begin{verbatim}
SELECT ...
FROM cast_info AS ci,
     company_name AS cn,
     keyword AS k,
     movie_companies AS mc,
     movie_keyword AS mk,
     name AS n,
     title AS t
WHERE k.keyword ='character-name-in-title'
  AND n.name LIKE 'X%'
  AND n.id = ci.person_id
  AND ci.movie_id = t.id
  AND t.id = mk.movie_id
  AND mk.keyword_id = k.id
  AND t.id = mc.movie_id
  AND mc.company_id = cn.id
  AND ci.movie_id = mc.movie_id
  AND ci.movie_id = mk.movie_id
  AND mc.movie_id = mk.movie_id;
\end{verbatim}

    Re-optimized Query
\begin{verbatim}
CREATE TEMP TABLE temp1 AS 
SELECT mk.movie_id
FROM keyword AS k,
     movie_keyword AS mk
WHERE mk.keyword_id = k.id
  AND k.keyword ='character-name-in-title';
  
SELECT ...
FROM cast_info AS ci,
     company_name AS cn,
     movie_companies AS mc,
     name AS n,
     title AS t,
     temp1 
WHERE n.name LIKE 'X%'
  AND n.id = ci.person_id
  AND ci.movie_id = t.id
  AND t.id = temp1.movie_id
  AND t.id = mc.movie_id
  AND mc.company_id = cn.id
  AND ci.movie_id = mc.movie_id
  AND ci.movie_id = temp1.movie_id
  AND mc.movie_id = temp1.movie_id;
\end{verbatim}
    \caption{
        Example of re-optimization simulation
    }
    \label{fig:re_opt_example}
\end{figure} 

\begin{figure*}[t]
    \centering     
    \includegraphics[width=\textwidth]{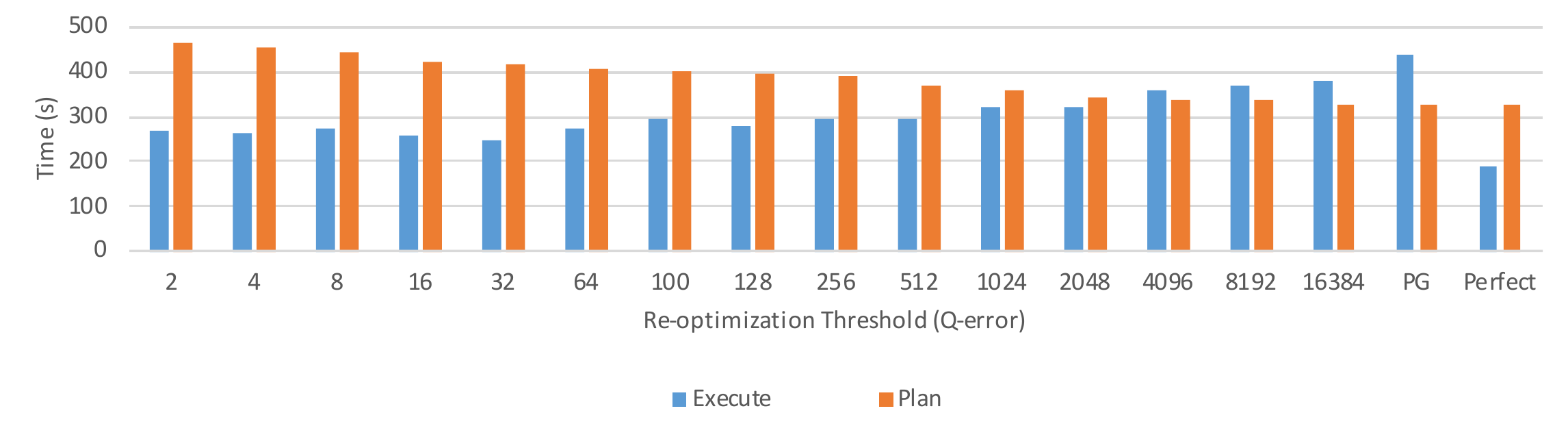}
    \caption{
        Total execution and planning time of all queries in \job. Comparing different re-optimization thresholds, \pg, and perfect-(17)
    }
    \label{fig:threshold_sweep}
\end{figure*} 

\begin{figure*}[t]
    \centering     
    \includegraphics[width=\textwidth]{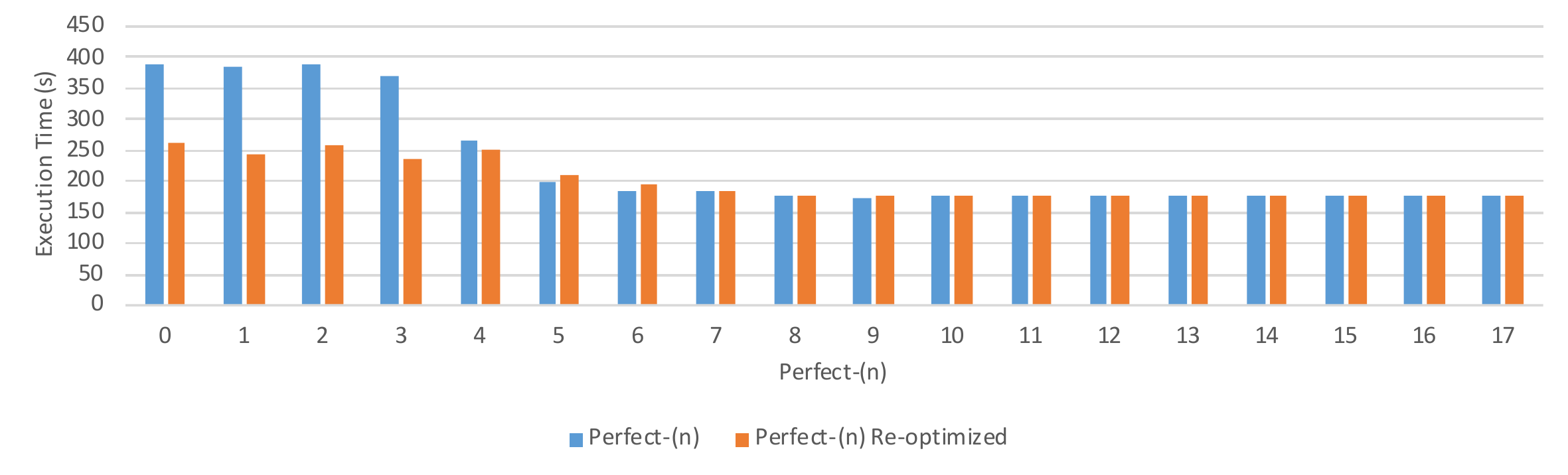}
    \caption{
        Total execution time of \job varying the quality of estimates from perfect-(0) to perfect-(17), with and without re-optimization
    }
    \label{fig:reopt_vs_perfect}
\end{figure*}

We simulate a simple query re-optimization scheme by examining the \texttt{EXPLAIN ANALYZE} output 
of each 
query and comparing the true cardinalities to the those predicted by the \pg Cardinality Estimator 
(CE). For the lowest join operator in the query plan with a true cardinality a factor of $n$ larger 
or smaller than the estimated cardinality, we rewrite this sub-query to create a temporary table 
instead. For the remainder of the query, we replace all the tables in the poorly estimated join 
with the temporary table and re-plan. We repeat this procedure until no join operators in the query 
plan have a cardinality estimate a factor of $n$ different from the true cardinality. 
While materialization of intermediate results may be expensive, it gives  the optimizer a chance to correct itself for the rest of the execution.

We then measure the planning and execution time for the resulting ``re-optimized'' query. For the 
planning time, we sum the planing time of the original query and the planning time of the 
\texttt{SELECT} queries. We do not include the time required to plan the creation of 
temporary tables since this is already included in the original query planning time. Thus the re-optimized query is a series of \texttt{CREATE TEMPORARY TABLE} commands followed by a \texttt{SELECT} query to generate the final result. An example of this re-optimization is shown in Figure~\ref{fig:re_opt_example}. 

Total execution time is measured by summing the execution time of each \texttt{CREATE TEMPORARY TABLE} command and the final \texttt{SELECT} query. We use execution and planning time as reported by
\pg's \texttt{EXPLAIN ANALYZE} command. Unless otherwise noted, we report only the query execution time and exclude planning time. For many queries short-running in the benchmark, \pg planning time exceeds execution time. Clearly it does not pay off to re-optimize these queries.

We believe this is a reasonable approximation of a simplistic re-optimization scheme. Although it is possible that this approximation breaks a pipeline present in the original query plan, our simulation provides a reasonable approximation for the upper bound of the cost of re-optimization schemes, since it requires a full materialization of intermediate tables. More sophisticated systems implementing re-optimization may perform better than our simulation.  

More complex re-optimization schemes have been proposed. Rio~\cite{babu2005proactive}, for instance, makes uncertainty explicit in cardinality estimates and creates execution plans that will automatically change when the true cardinality lies outside a range of predicted cardinalities with a single planning phase. We believe that using advanced optimization mechanisms like Rio may achieve better end-to-end latency then we report with our simulation since they avoid some of the costs of stopping the query to re-plan.

\subsection{Re-optimization Triggers}

Clearly the decision of when to re-optimize is critical to a scheme's total performance. In our setup 
we re-optimize when the relative cardinality estimate error crosses a threshold. If the expected cardinality of a join is, n times more or less than we expect, we materialize the result, and re-optimize the rest of the query. That is, we re-optimize a query when the Q-error~\cite{moerkotte2009preventing} exceeds a threshold value. If the threshold is too low, even good plans may be unnecessarily re-optimized. 
If set too high, re-optimization will never be triggered. In Figure~\ref{fig:threshold_sweep} we 
vary the re-optimization threshold and observe the impact on planning and execution time, comparing to \pg and using perfect estimates. Based on these experiments, we find that setting the threshold to 32 gives the best query execution time improvement. Therefore we set the threshold to 32 for the following experiments.

While we chose to re-optimize only by comparing the predicted cardinality to the actual cardinality 
of each of the join operators. This is not the only possible trigger for re-optimization. Prior work in adaptive query processing can change query execution based on several sources of feedback. Cuttlefish, for instance, can use tuple processing time of an operator to choose a different implementation at runtime using reinforcement learning techniques~\cite{kaftan2018cuttlefish}.

Figure~\ref{fig:threshold_sweep} shows the impact of different thresholds on query planning and execution time. Surprisingly, even re-optimizing at a threshold of 2, the lowest we tested, only increases the query planning time by about 42\% over \pg while decreasing the execution time by 40\%. It is worth noting that the planning time remains constant independent of the data size while the execution time will increase with the size of the data. While Figure~\ref{fig:threshold_sweep}, shows that planning time is significant, sometimes exceeding execution time, this is a result of the small size of the IMDB dataset, rather than the excessive costs of planning. 

Also surprising is that frequent materialization of tables by setting the threshold to a relatively low value of two has a minimal impact on query execution time. Setting the re-optimization threshold to two only degrades performance 10\% compared to the lowest re-optimization execution time measured. This indicates that, at least for \job, setting the re-optimization threshold too low is still better than not re-optimizing at all. With longer running queries, the additional cost of planning queries is likely negligible. Because we expect that re-optimization will provide the most benefit for longer running queries taking several minutes, we report only execution time for the following experiments.

\begin{table}
    \centering
    {\scriptsize {
\setlength{\tabcolsep}{3pt}
\begin{tabular}{l|l}
	\hline
        \textbf{Relative Runtime} & \textbf{Number of Queries} \\
        \hline
    0.1 - 0.8            &  6 \\
    0.8 - 1.2            &  47 \\
    1.2 - 2.0            &  21 \\
    2.0 - 5.0            &  29 \\
    $>$ 5.0            &  10 \\ 
\end{tabular}
}
}
    \vspace{5px}
    \caption{
        Execution time of \job queries with re-optimization relative to perfect-(17)
    }
    \label{tab:re-opt}
\end{table}

\subsection{Re-optimization and Better Cardinality Estimates}
\label{sec:reopt_and_card}

As we discuss in Section~\ref{sec:background}, we believe that without a redesign of the 
query optimizer, query re-optimization is most likely to 
lead to the biggest gains in end-to-end query latency. However, if a method to improve cardinalities approaches perfect-(3) or perfect-(4), re-optimization can still improve execution time. As cardinality estimates improve, the need to re-optimize decreases. In 
Figure~\ref{fig:reopt_vs_perfect}, we compare perfect-(n) plus re-optimization for varying values of n. We see that 
re-optimization improves the latency of perfect-(n) estimates until perfect-(5). While re-optimizing perfect-(5) slows the execution of the workload, the risk is relatively small, The execution time of the benchmark is only 6\% slower with re-optimization than only perfect-(5).

\subsection{The Benefits of Re-optimization}

\begin{figure*}[t]
    \centering     
    \includegraphics[width=\textwidth]{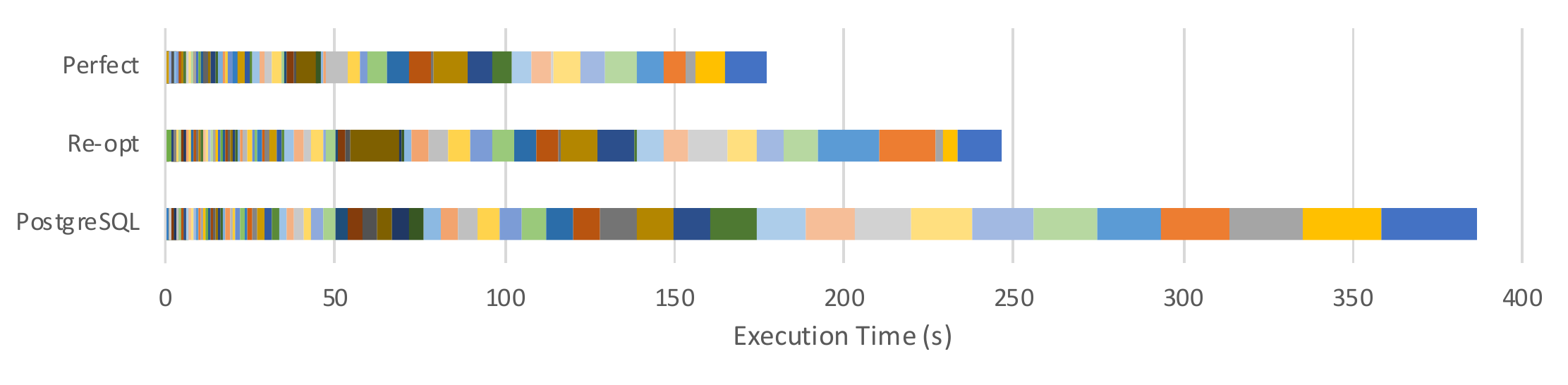}
    \caption{
        Total execution time comparison of perfect, re-optimized, and \pg by query, ordered by execution time on \pg.
    }
    \label{fig:query_execution_time_plot}
\end{figure*}

In Figure~\ref{fig:query_execution_time_plot} we compare the execution time of standard \pg to perfect-(17) and our re-optimization simulation. We visualize the total execution time by query, and order queries from shortest to longest in the original \pg execution. Note that this graph does not include planning time. While re-optimization does not significantly improve the execution time of most of the shortest queries, the impact of re-optimization on the longest running queries achieves much of the benefits of having perfect estimates. Thus, we find there is no need to re-optimize the shortest queries, particularly because the re-optimization may exceed query execution time.

Table~\ref{tab:re-opt} shows the distribution of the execution time of queries after re-optimization. This table demonstrates that many more queries are performing close to optimally than before re-optimization. We ignore planning time since we assume that most queries come from an application and will have cached plans. Figure~\ref{fig:query_execution_time_plot} and Figure~\ref{fig:top_20_comparison} reinforce this. Many of the longest running queries in the benchmark have marked improvement after re-optimizing, improving the execution time of the whole benchmark by 45\% over \pg, and achieving more than half of the benefit of perfect estimates, at the cost of longer re-planning times. 

\subsection{The Risks of Re-optimization}

While query re-optimization has a clear benefit for the \job it is not without risks. Figure~\ref{fig:query_execution_time_plot} shows the execution time of the benchmark by query. It compares default \pg, re-optimization, and perfect-(17) ordered by the default \pg execution time. In several queries the execution time increases significantly from default \pg. In fact, the worst performing re-optimized query has more than 100 times worse execution time than the original query executed by \pg. However, since the original execution time of the query is only about ten milliseconds, the impact on the execution time of the whole benchmark is negligible. This can be avoided by re-optimizing only long-running queries.

\section{Conclusion}
\label{sec:conclusion}

In this paper we showed that a simple re-optimization strategy achieves much of the benefit of perfect cardinality estimation in the Join Order Benchmark. While we confirm that having perfect cardinality estimates is a clear way to improve query plans, we show that achieving estimates good enough to impact execution time of a workload is daunting.

Our results show the promise of re-optimization for a single workload with a small dataset with all data and indexes cached. Given these results we believe that more investigation into re-optimization strategies is needed. Much of re-optimization work was performed more than a decade ago. Our experiments show that re-optimization significantly improves query execution time compared to default \pg cardinality estimates on a single threaded row store. However, modern query execution engines are heavily pipelined, compile queries, execute on many machines, or use columnar storage. Re-optimization breaks this pipeline. And if a query is re-planned, this may require another expensive compilation phase. Some of this cost might be mitigated by adopting the strategy similar to Rio~\cite{babu2005proactive}, where the query plan has uncertainty built-in. But the costs of doing so are still unclear. More work is needed to determine the feasibility of re-optimization for modern workloads on modern systems.

\bibliographystyle{IEEEtran}
\bibliography{card_est.bib}

\begin{thebibliography}{10}
\providecommand{\url}[1]{#1}
\csname url@samestyle\endcsname
\providecommand{\newblock}{\relax}
\providecommand{\bibinfo}[2]{#2}
\providecommand{\BIBentrySTDinterwordspacing}{\spaceskip=0pt\relax}
\providecommand{\BIBentryALTinterwordstretchfactor}{4}
\providecommand{\BIBentryALTinterwordspacing}{\spaceskip=\fontdimen2\font plus
\BIBentryALTinterwordstretchfactor\fontdimen3\font minus
  \fontdimen4\font\relax}
\providecommand{\BIBforeignlanguage}[2]{{%
\expandafter\ifx\csname l@#1\endcsname\relax
\typeout{** WARNING: IEEEtran.bst: No hyphenation pattern has been}%
\typeout{** loaded for the language `#1'. Using the pattern for}%
\typeout{** the default language instead.}%
\else
\language=\csname l@#1\endcsname
\fi
#2}}
\providecommand{\BIBdecl}{\relax}
\BIBdecl

\bibitem{selinger1979access}
P.~G. Selinger, M.~M. Astrahan, D.~D. Chamberlin, R.~A. Lorie, and T.~G. Price,
  ``Access path selection in a relational database management system,'' in
  \emph{Proceedings of the 1979 ACM SIGMOD international conference on
  Management of data}.\hskip 1em plus 0.5em minus 0.4em\relax ACM, 1979, pp.
  23--34.

\bibitem{estan2006end}
C.~Estan and J.~F. Naughton, ``End-biased samples for join cardinality
  estimation,'' in \emph{Data Engineering, 2006. ICDE'06. Proceedings of the
  22nd International Conference on}.\hskip 1em plus 0.5em minus 0.4em\relax
  IEEE, 2006, pp. 20--20.

\bibitem{chen2017two}
Y.~Chen and K.~Yi, ``Two-level sampling for join size estimation,'' in
  \emph{Proceedings of the 2017 ACM International Conference on Management of
  Data}.\hskip 1em plus 0.5em minus 0.4em\relax ACM, 2017, pp. 759--774.

\bibitem{leis2017cardinality}
V.~Leis, B.~Radke, A.~Gubichev, A.~Kemper, and T.~Neumann, ``Cardinality
  estimation done right: Index-based join sampling.'' in \emph{CIDR}, 2017.

\bibitem{marcus2018deep}
R.~Marcus and O.~Papaemmanouil, ``Deep reinforcement learning for join order
  enumeration,'' in \emph{Proceedings of the First International Workshop on
  Exploiting Artificial Intelligence Techniques for Data Management}.\hskip 1em
  plus 0.5em minus 0.4em\relax ACM, 2018, p.~3.

\bibitem{ortiz2018learning}
J.~Ortiz, M.~Balazinska, J.~Gehrke, and S.~S. Keerthi, ``Learning state
  representations for query optimization with deep reinforcement learning,'' in
  \emph{Proceedings of the Second Workshop on Data Management for End-To-End
  Machine Learning}.\hskip 1em plus 0.5em minus 0.4em\relax ACM, 2018, p.~4.

\bibitem{kabra1998efficient}
N.~Kabra and D.~J. DeWitt, ``Efficient mid-query re-optimization of sub-optimal
  query execution plans,'' in \emph{ACM SIGMOD Record}, vol.~27, no.~2.\hskip
  1em plus 0.5em minus 0.4em\relax ACM, 1998, pp. 106--117.

\bibitem{babu2005proactive}
S.~Babu, P.~Bizarro, and D.~DeWitt, ``Proactive re-optimization,'' in
  \emph{Proceedings of the 2005 ACM SIGMOD international conference on
  Management of data}.\hskip 1em plus 0.5em minus 0.4em\relax ACM, 2005, pp.
  107--118.

\bibitem{kaftan2018cuttlefish}
T.~Kaftan, M.~Balazinska, A.~Cheung, and J.~Gehrke, ``Cuttlefish: A lightweight
  primitive for adaptive query processing,'' \emph{arXiv preprint
  arXiv:1802.09180}, 2018.

\bibitem{avnur2000eddies}
R.~Avnur and J.~M. Hellerstein, ``Eddies: Continuously adaptive query
  processing,'' in \emph{ACM sigmod record}, vol.~29, no.~2.\hskip 1em plus
  0.5em minus 0.4em\relax ACM, 2000, pp. 261--272.

\bibitem{deshpande2007adaptive}
A.~Deshpande, Z.~Ives, V.~Raman \emph{et~al.}, ``Adaptive query processing,''
  \emph{Foundations and Trends{\textregistered} in Databases}, vol.~1, no.~1,
  pp. 1--140, 2007.

\bibitem{leis2015good}
V.~Leis, A.~Gubichev, A.~Mirchev, P.~Boncz, A.~Kemper, and T.~Neumann, ``How
  good are query optimizers, really?'' \emph{Proceedings of the VLDB
  Endowment}, vol.~9, no.~3, pp. 204--215, 2015.

\bibitem{babcock2005towards}
B.~Babcock and S.~Chaudhuri, ``Towards a robust query optimizer: a principled
  and practical approach,'' in \emph{Proceedings of the 2005 ACM SIGMOD
  international conference on Management of data}.\hskip 1em plus 0.5em minus
  0.4em\relax ACM, 2005, pp. 119--130.

\bibitem{manegold2002generic}
S.~Manegold, P.~Boncz, and M.~L. Kersten, ``Generic database cost models for
  hierarchical memory systems,'' in \emph{Proceedings of the 28th international
  conference on Very Large Data Bases}.\hskip 1em plus 0.5em minus 0.4em\relax
  VLDB Endowment, 2002, pp. 191--202.

\bibitem{kipf2018learned}
A.~Kipf, T.~Kipf, B.~Radke, V.~Leis, P.~Boncz, and A.~Kemper, ``Learned
  cardinalities: Estimating correlated joins with deep learning,'' \emph{arXiv
  preprint arXiv:1809.00677}, 2018.

\bibitem{krishnan2018learning}
S.~Krishnan, Z.~Yang, K.~Goldberg, J.~Hellerstein, and I.~Stoica, ``Learning to
  optimize join queries with deep reinforcement learning,'' \emph{arXiv
  preprint arXiv:1808.03196}, 2018.

\bibitem{ibaraki1984optimal}
T.~Ibaraki and T.~Kameda, ``On the optimal nesting order for computing
  n-relational joins,'' \emph{ACM Transactions on Database Systems (TODS)},
  vol.~9, no.~3, pp. 482--502, 1984.

\bibitem{graefe1993volcano}
G.~Graefe and W.~McKenna, ``The volcano optimizer generator: extensibility and
  efficient search,'' in \emph{Data Engineering, 1993. Proceedings. Ninth
  International Conference on}.\hskip 1em plus 0.5em minus 0.4em\relax IEEE,
  1993, pp. 209--218.

\bibitem{chaudhuri1998overview}
S.~Chaudhuri, ``An overview of query optimization in relational systems,'' in
  \emph{Proceedings of the seventeenth ACM SIGACT-SIGMOD-SIGART symposium on
  Principles of database systems}.\hskip 1em plus 0.5em minus 0.4em\relax ACM,
  1998, pp. 34--43.

\bibitem{graefe1995cascades}
G.~Graefe, ``The cascades framework for query optimization,'' \emph{IEEE Data
  Eng. Bull.}, vol.~18, no.~3, pp. 19--29, 1995.

\bibitem{chen2016memsql}
J.~Chen, S.~Jindel, R.~Walzer, R.~Sen, N.~Jimsheleishvilli, and M.~Andrews,
  ``The memsql query optimizer: A modern optimizer for real-time analytics in a
  distributed database,'' \emph{Proceedings of the VLDB Endowment}, vol.~9,
  no.~13, pp. 1401--1412, 2016.

\bibitem{postgres}
T.~P. G.~D. Group, \emph{Documentation PostgreSQL 10.1}, 2018.

\bibitem{bennett1991genetic}
K.~Bennett, M.~C. Ferris, and Y.~E. Ioannidis, \emph{A genetic algorithm for
  database query optimization}.\hskip 1em plus 0.5em minus 0.4em\relax Computer
  Sciences Department, University of Wisconsin, Center for Parallel
  Optimization, 1991.

\bibitem{waas2000join}
F.~Waas and A.~Pellenkoft, ``Join order selection (good enough is easy),'' in
  \emph{British National Conference on Databases}.\hskip 1em plus 0.5em minus
  0.4em\relax Springer, 2000, pp. 51--67.

\bibitem{ioannidis1993optimal}
Y.~E. Ioannidis and S.~Christodoulakis, ``Optimal histograms for limiting
  worst-case error propagation in the size of join results,'' \emph{ACM
  Transactions on Database Systems (TODS)}, vol.~18, no.~4, pp. 709--748, 1993.

\bibitem{ioannidis1995balancing}
Y.~E. Ioannidis and V.~Poosala, ``Balancing histogram optimality and
  practicality for query result size estimation,'' in \emph{Acm Sigmod Record},
  vol.~24, no.~2.\hskip 1em plus 0.5em minus 0.4em\relax ACM, 1995, pp.
  233--244.

\bibitem{poosala1996improved}
V.~Poosala, P.~J. Haas, Y.~E. Ioannidis, and E.~J. Shekita, ``Improved
  histograms for selectivity estimation of range predicates,'' in \emph{ACM
  Sigmod Record}, vol.~25, no.~2.\hskip 1em plus 0.5em minus 0.4em\relax ACM,
  1996, pp. 294--305.

\bibitem{jagadish1998optimal}
H.~V. Jagadish, N.~Koudas, S.~Muthukrishnan, V.~Poosala, K.~C. Sevcik, and
  T.~Suel, ``Optimal histograms with quality guarantees,'' in \emph{VLDB},
  vol.~98, 1998, pp. 24--27.

\bibitem{piatetsky1984accurate}
G.~Piatetsky-Shapiro and C.~Connell, ``Accurate estimation of the number of
  tuples satisfying a condition,'' \emph{ACM Sigmod Record}, vol.~14, no.~2,
  pp. 256--276, 1984.

\bibitem{ioannidis1991propagation}
Y.~E. Ioannidis and S.~Christodoulakis, \emph{On the propagation of errors in
  the size of join results}.\hskip 1em plus 0.5em minus 0.4em\relax ACM, 1991,
  vol.~20, no.~2.

\bibitem{alon2002tracking}
N.~Alon, P.~B. Gibbons, Y.~Matias, and M.~Szegedy, ``Tracking join and
  self-join sizes in limited storage,'' \emph{Journal of Computer and System
  Sciences}, vol.~64, no.~3, pp. 719--747, 2002.

\bibitem{ilyas2004cords}
I.~F. Ilyas, V.~Markl, P.~Haas, P.~Brown, and A.~Aboulnaga, ``Cords: automatic
  discovery of correlations and soft functional dependencies,'' in
  \emph{Proceedings of the 2004 ACM SIGMOD international conference on
  Management of data}.\hskip 1em plus 0.5em minus 0.4em\relax ACM, 2004, pp.
  647--658.

\bibitem{haas1999ripple}
P.~J. Haas and J.~M. Hellerstein, ``Ripple joins for online aggregation,''
  \emph{ACM SIGMOD Record}, vol.~28, no.~2, pp. 287--298, 1999.

\bibitem{tpc-h}
{The Transaction Processing Council}, ``{TPC-H} {B}enchmark ({R}evision
  2.16.0),'' \url{http://www.tpc.org/tpch/}, June 2013.

\bibitem{stillger2001leo}
M.~Stillger, G.~M. Lohman, V.~Markl, and M.~Kandil, ``Leo-db2's learning
  optimizer,'' in \emph{VLDB}, vol.~1, 2001, pp. 19--28.

\bibitem{moerkotte2009preventing}
G.~Moerkotte, T.~Neumann, and G.~Steidl, ``Preventing bad plans by bounding the
  impact of cardinality estimation errors,'' \emph{Proceedings of the VLDB
  Endowment}, vol.~2, no.~1, pp. 982--993, 2009.

\end{thebibliography}

\end{document}